\documentclass[sigconf]{acmart}

\AtBeginDocument{%
  \providecommand\BibTeX{{%
    \normalfont B\kern-0.5em{\scshape i\kern-0.25em b}\kern-0.8em\TeX}}}

\copyrightyear{2025}
\acmYear{2025}
\setcopyright{rightsretained}
\acmConference[WSDM '25]{Proceedings of the Eighteenth ACM International Conference on Web Search and Data Mining}{March 10--14, 2025}{Hannover, Germany}
\acmBooktitle{Proceedings of the Eighteenth ACM International Conference on Web Search and Data Mining (WSDM '25), March 10--14, 2025, Hannover, Germany}\acmDOI{10.1145/3701551.3703507}
\acmISBN{979-8-4007-1329-3/25/03}

\usepackage{multirow}
\usepackage{graphicx}
\usepackage{epstopdf}
\usepackage{subcaption}
\usepackage{bbding}
\usepackage{enumitem}
\usepackage{balance}
\usepackage[linesnumbered,ruled,vlined]{algorithm2e}
\usepackage{hyperref}
\makeatletter
\gdef\@copyrightpermission{
 \begin{minipage}{0.3\columnwidth}
 \href{https://creativecommons.org/licenses/by/4.0/}{\includegraphics[width=0.90\textwidth]{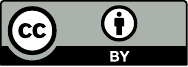}}
 \end{minipage}\hfill
 \begin{minipage}{0.7\columnwidth}
 \href{https://creativecommons.org/licenses/by/4.0/}{This work is licensed under a Creative Commons
Attribution International 4.0 License.}
 \end{minipage}
 \vspace{5pt}
}
\makeatother

\begin{document}

\title[Teach Me How to Denoise: A Universal Framework for Denoising Multi-modal Recommender \\ Systems via Guided Calibration]{Teach Me How to Denoise: A Universal Framework for Denoising Multi-modal Recommender Systems via Guided Calibration}


\author{Hongji Li}
\authornote{Equal contribution.}
\affiliation{%
  \institution{Lanzhou University}
  \city{Lanzhou}
  \country{China}}
\email{3420670269neon@gmail.com}

\author{Hanwen Du}
\authornotemark[1]
\affiliation{%
  \institution{The Ohio State University}
  \city{Columbus}
  \country{USA}}
\email{du.1128@osu.edu}

\author{Youhua Li}
\affiliation{%
  \institution{City University of Hong Kong}
  \city{Hong Kong}
  \country{China}}
\email{liyouhua97@gmail.com}

\author{Junchen Fu}
\affiliation{%
  \institution{University of Glasgow}
  \city{Glasgow}
  \country{UK}}
\email{j.fu.3@research.gla.ac.uk}

\author{Chunxiao Li}
\affiliation{%
  \institution{University of Science and Technology of China}
  \city{Hefei}
  \country{China}}
\email{chunxiao.li@ustc.edu.cn}

\author{Ziyi Zhuang}
\affiliation{%
  \institution{Shanghai Jiao Tong University}
  \city{Shanghai}
  \country{China}}
\email{ziyi123@sjtu.edu.cn}

\author{Jiakang Li}
\affiliation{%
  \institution{Rutgers University}
  \city{New Brunswick}
  \country{USA}}
\email{jiakang.li@rutgers.edu}

\author{Yongxin Ni}
\authornote{Corresponding author.}
\affiliation{%
  \institution{National University of Singapore}
  \city{Singapore}
  \country{Singapore}}
\email{niyongxin@u.nus.edu }

\renewcommand{\shortauthors}{Hongji Li, et al.}

\begin{abstract}
   The surge in multimedia content has led to the development of Multi-Modal Recommender Systems (MMRecs), which use diverse modalities—like text, images, videos, and audio—for more personalized recommendations. However, MMRecs struggle with the challenge of noisy data from the misalignment among modal content and the difference between modal and recommendation semantics, while traditional denoising methods fall short in addressing these issues due to the complexity of multi-modal data. To overcome this, we propose a universal \underline{gu}ided \underline{i}n-sync \underline{d}istillation d\underline{e}noising framework for multi-modal \underline{r}ecommndation (GUIDER), aimed at improving MMRecs by denoising user feedbacks. Specifically, GUIDER employs a re-calibration strategy to identify clean and noisy interactions from modal content. Furthermore, it incorporates a Denoising Bayesian Personalized Ranking (DBPR) loss function to denoise implicit user feedback. Finally, it utilizes a denoising knowledge distillation objective based on Optimal Transport (OT) distance to guide the mapping from modality representations to recommendation semantics spaces. GUIDER can be seamlessly integrated into existing MMRecs methods as a plug-and-play solution for recommendation denoising. Experiment results on four public datasets show its effectiveness and universality across various MMRecs.
   Our source code is openly available on GitHub at \url{https://github.com/Neon-Jing/Guider}
   
\end{abstract}




\maketitle
\begin{figure}[htbp]
  \centering
   \includegraphics[width=\linewidth]{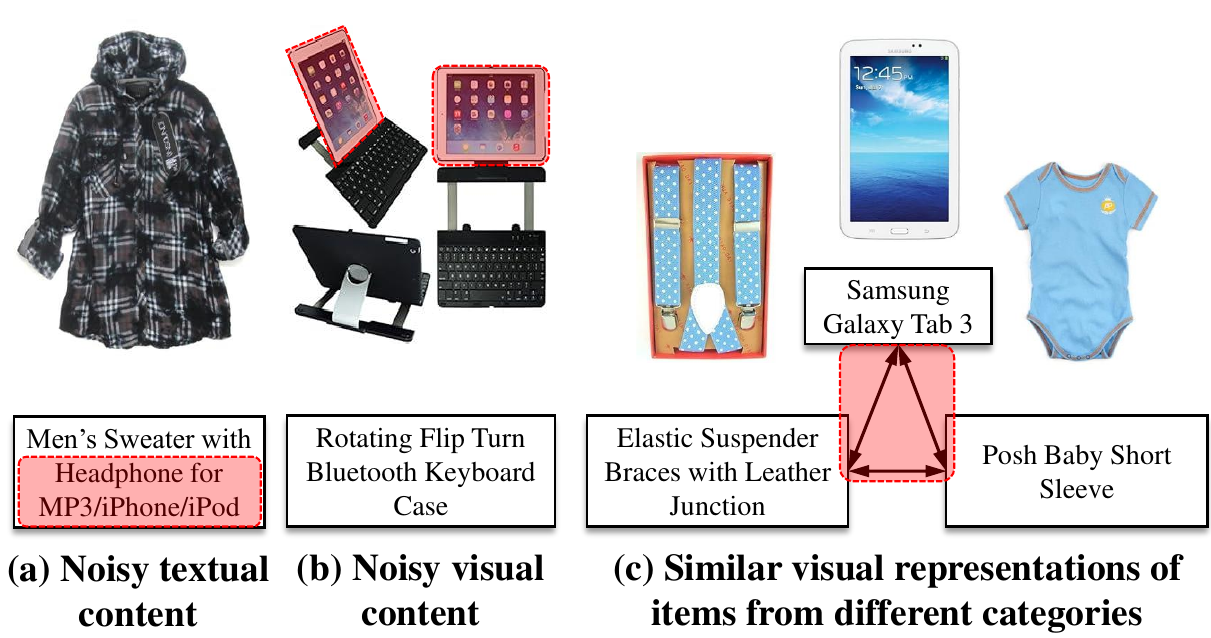}
  \caption{An illustration of challenges in denoising MMRecs.}
    \label{noise} 
\end{figure}
\section{Introduction}
Recommender systems serve as powerful information filtering engines with widespread applications in various scenarios, such as e-Commerce \cite{schafer2001commerce}, short video sharing \cite{wei2019mmgcn,ni2023content}, and social media platforms \cite{guy2010social}. Traditional recommender systems are ID-based \cite{rendle2010factorization,he2017neural,Hidasi2016session}, where each user/item is assigned a unique identifier (ID) and subsequently converted into embeddings to learn preferences from user-item interactions. To leverage the rich sources of multimedia content to explore user interest, Multi-Modal Recommender Systems (MMRecs) \cite{chen2017attentive,wei2019mmgcn,wei2023multi,yuan2023go,ji2023online,zhang2024ninerec,cheng2024image} incorporate the multi-modal contents of items (e.g., product text captions and image covers) via model-specific encoders and mine the fine-grained relations between modalities to capture modal-specific user preferences. Such a paradigm shift can alleviate the cold-start issue \cite{hou2022towards} and shows better transferability \cite{wang2022beibei,yuan2023go,li2024multi,fu2024exploring,fu2024iisan,fu2024efficient,li2024empirical}, and is further popularized by the advent of powerful multi-modal foundation models such as GPT-4 \cite{achiam2023gpt}, CLIP \cite{radford2021learning}, DALLE \cite{ramesh2022hierarchical} and LLaMa \cite{touvron2023llama}. 

A formidable obstacle to developing high-performing recommender systems lies in the \emph{noisy interactions} issue. Similar to ID-based Recommender Systems (IDRecs), MMRecs also suffer from noisy interactions introduced by spurious user feedback, which can stem from accidental clicks or curiosity-driven behaviors that do not accurately reflect user preferences. Furthermore, denoising under the MMRecs setting faces unique challenges due to the inclusion of multi-modal information, as depicted in Figure~\ref{noise}. The first challenge is the noise from modal content, where text or image descriptions can be misleading or irrelevant (Figure~\ref{noise}(a) and 2(b)). The second challenge is mapping inconsistencies, where visual representations of items from very different categories (e.g., suspenders, smartphones, and baby clothing) appear similar, leading to semantic misalignment in recommendations (Figure~\ref{noise}(c)).

\begin{figure}
    \centering
\includegraphics[width=\linewidth]{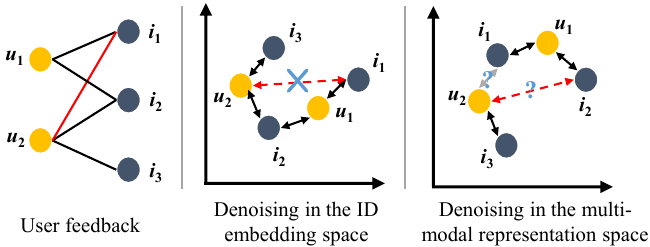}
    \caption{The challenges of denoising in the multi-modal representation space. 
    }
    \label{denoiseillustration}
\end{figure}

To alleviate the noisy interaction issue, most existing works focus on ID-based methods and mainly propose to denoise implicit feedback through \emph{sample selection} \cite{wang2021denoising,tian2022learning} or \emph{sample reweighting} \cite{wang2021denoising,gantner2012personalized, gao2022self}. The intuition behind these approaches is that noisy user-item interactions are relatively hard to fit into a recommender model and, therefore, tend to exhibit lower similarity scores and contribute to larger loss values. However, our experiments in Section~\ref{empiricalstudies} reveal that this assumption, while valid in the IDRec setting, breaks down in the MMRecs setting due to the significant variance between the ID embedding space and the multi-modal representation space, as illustrated in Figure~\ref{denoiseillustration}. In the multi-modal space, 
noisy user-item interactions (e.g., $(u_2,i_1)$) exhibiting lower similarity score in the ID embedding space are no longer detectable, and true user-item interactions (e.g., $(u_2,i_2)$) with noisy modal representations may also exhibit large loss value, making it difficult to detect noisy interactions. This discrepancy leads to the misidentification of many ground-truth user-item interactions as noise when applying traditional denoising losses directly to MMRecs. Therefore, we argue that methods designed for IDRec are inadequate for denoising MMRecs. It is crucial to develop a general and effective denoising framework tailored to the unique properties of MMRecs.

To tackle the challenges mentioned above, we propose a guided in-sync distillation denoising framework for multi-modal recommendation (GUIDER), a universal denoising framework designed to enhance MMRecs. Specifically, GUIDER introduces Adaptive Modality Similarity Calibration (AMSC) to identify noisy interactions from modal content by redistributing reliable and spurious samples via fine-grained re-calibration strategies based on the semantic similarity between text and vision modalities. Next, GUIDER introduces a Denoising Bayesian Personalized Ranking (DBPR) loss function to tackle noisy user feedbacks through the positive and noisy negatives identified by AMSC. While our empirical studies show that it is infeasible to directly apply denoising loss functions to MMRecs, we bypass this issue by training a teacher IDRec with DBPR and then employing the denoised teacher to guide the training of the MMRec as a student via Knowledge Distillation (KD) \cite{hinton2015distilling}, thereby effectively removing the noises during the transition process from the modal representation space to the recommendation semantic space. Moreover, as the standard KL-divergence may suffer from over-smoothing \cite{kim2016sequence} and model-collapse \cite{arjovsky2017towards}, we propose a novel Optimal Transport (OT) distance \cite{cuturi2013sinkhorn} based KD objective for distilling recommendation logits, which is more suitable to be a symmetric distance metric for KD with stable gradient \cite{shen2018wasserstein}. 

Our contributions can be summarized as follows:
\begin{itemize}[leftmargin=8pt,topsep=1pt]
    \item We identify the unique challenges faced with denoising MMRecs and reveal the limitations of existing IDRec denoising methods in handling multi-modal noises.
    \item In light of these challenges, we propose GUIDER, a universal denoising framework for MMRecs with via fine-grained re-calibration strategy from modalities semantics and guided KD based on OT distance, performing effective denoising from multi-modal content and significantly improving recommendation performances.
    \item We conduct extensive experiments on four benchmark datasets to verify the effectiveness and universality of GUIDER, demonstrating its versatility as a plug-and-play denoising framework for MMRecs across diverse datasets and noise conditions.
\end{itemize}

\section{Related Works}
\subsection{Multi-modal Recommender System}
Multi-modal recommender systems leverage diverse multimedia sources to enhance recommendation accuracy by integrating information from various modalities within the collaborative filtering framework. For example, VBPR \cite{he2016vbpr} extends the traditional BPR method \cite{rendle2012bpr} by incorporating visual features, VECF \cite{chen2019personalized} performs pre-segmentation on images, utilizing visual features extracted from product images and capturing user attention on different image regions. 
In the trend of applying Graph Neural Networks (GNN) to recommendation systems \cite{wu2022graph}, MMGCN \cite{wei2019mmgcn} constructs a modality-specific user-item bipartite graph and a message-passing mechanism to capture information from multi-hop neighbors, GRCN \cite{wei2020graph} introduces a graph refinement layer to refine the structure of the user-item interaction graph, DRAGON \cite{zhou2023enhancing} enhances the dyadic relations by constructing homogeneous Graphs for multimodal recommendation. Recently, given the success of self-supervised representation learning \cite{he2020momentum, chen2020simple} in handling data sparsity, a trend of applying self-supervised learning methods to multi-modal recommender systems has emerged. For example, BM3 \cite{zhou2023bootstrap} proposes an efficient self-supervised learning framework for MMRecs by bootstraping latent contrastive views from the representations of users and items, MMGCL \cite{yi2022multi} introduces a graph-based self-supervised learning method into micro-video recommendation and introduces a novel negative sampling method to learn the correlations between modalities, FREEDOM \cite{zhou2023tale} introduces an MMRec by freezing the item-item graph and simultaneously denoising the user-item interaction graph through degree-sensitive edge pruning.
\subsection{Recommendation Denoising}
\begin{figure*}
    \centering
    \includegraphics[width = 1.0\linewidth]{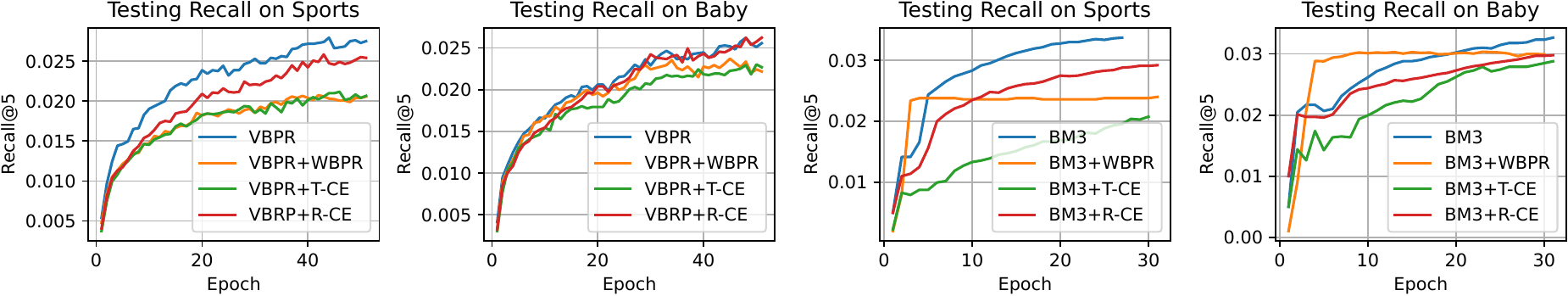}
    \caption{Training curves of different MMRecs equipped with different denoising methods on the Sports and Baby dataset.}
    \label{trainingcurve}
\end{figure*}
\begin{figure*}
    \centering
\includegraphics[width = 1.0\linewidth]{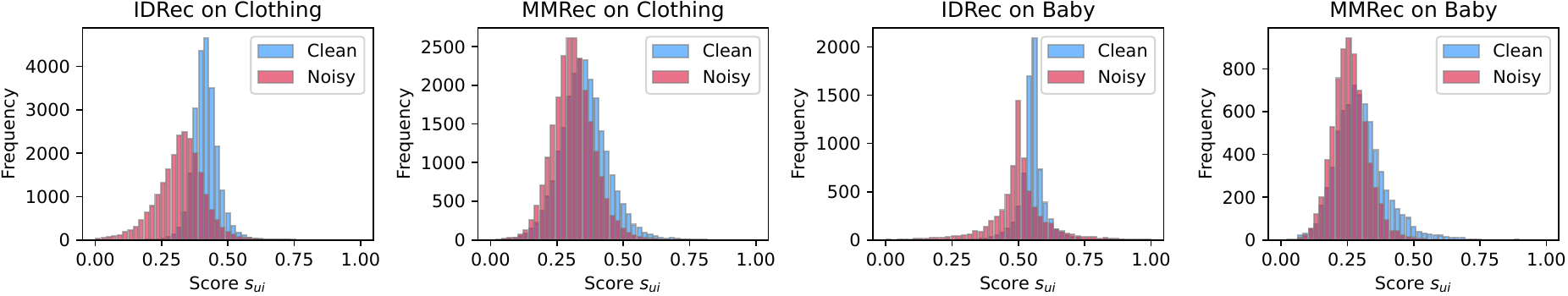}
    \caption{Normalized similarity score distributions of IDRec and MMRec on the Clothing and Baby dataset.}
    \label{scoredistribution}
\end{figure*}
The ubiquity of noises within recommendation system datasets necessitates the need to purify the user-item interactions via denoising methods. The classical WBPR \cite{gantner2012personalized} method considers popular items not yet interacted with by the user as true negative feedback. T-CE \cite{wang2021denoising} adjusts the loss function to truncate the contribution of noisy interactions, identified by their loss values exceeding a dynamically adjusted threshold. R-CE \cite{wang2021denoising} applies a re-weighting function to the loss values to mitigate the influence of noisy interactions.
SGDL \cite{wu2021self} utilizes the model memorization effect to guide recommendation denoising and introduces a denoising scheduler to
 improve the robustness.  
DeCA \cite{wang2022learning} utilizes the agreement across different model predictions as the denoising signal for implicit feedback. BOD \cite{wang2023efficient} proposes an efficient bi-level optimization framework to learn the denoising weights for user-item interactions.

\textbf{Remark.} 
Existing research on the multi-modal noise issues in MMRecs is still limited.  Our GUIDER addresses this gap by proposing a general framework that shifts the denoising task from traditional ID-based recommendation systems to MMRecs.
\section{Preliminaries}
\subsection{Problem Statement}
\noindent \textbf{IDRec and MMRecs.} We denote $\mathcal{U}$ as a set users, $\mathcal{I}$ as a set of items, and the user-item interaction data (e.g., user's implicit feedback such as click and review) $\mathcal{D}=\{u, i, r_{ui}|u\in\mathcal{U},i\in\mathcal{I}\}$ with the label $r_{ui}\in\{0, 1\}$ indicating whether user $u$ has interacted item $i$. 
To capture the user's preference from the implicit feedback, recommender systems usually learn the user representation $\{\mathbf{h}_{u}\in\mathbb{R}^{d}|u\in\mathcal{U}\}$ and the item representation $\{\mathbf{h}_{i}\in\mathbb{R}^{d}|i\in\mathcal{I}\}$ with the latent dimensionality $d$, and optimize the model with a recommendation loss (e.g., the pointwise BCE loss):
\begin{equation}
\label{recommendationloss}
\begin{aligned}
 &\mathcal{L}_{\text{Rec}}(\mathcal{D}) = \sum_{(u, i, r_{ui}) \in D}-r_{ui}\log(s_{ui})-(1-r_{ui})\log(1-s_{ui})\\
 \end{aligned}
\end{equation}
where $s_{ui}=\mathbf{h}^{\top}_{u}\cdot\mathbf{h}_{i}$ is the predicted similarity score between the user-item pair $(u,i)$. In traditional IDRec, $\mathbf{h}_{u}$ and $\mathbf{h}_{i}$ are derived from embedding matrices, while in MMRecs, $\mathbf{h}_{u}$ and $\mathbf{h}_{i}$ are typically obtained through fusing textual and visual representations extracted from modal-specific encoders.

\noindent \textbf{Denoise of IDRec.} As implicit feedback may not truly reflect user preference, previous researches \cite{gantner2012personalized,wang2021denoising} propose to denoise spurious user-item interactions by assigning a hard binary weight (sample selection) or a soft weight (sample reweighting) to $\mathcal{L}_{\text{Rec}}(\mathcal{D})$, resulting in a denoised recommendation loss $\mathcal{L}_{\text{denoise}}(u, i)=w_{ui}\mathcal{L}_{\text{Rec}}(u, i)$. To determine the appropriate value of $w_{ui}$, they assume that noisy user-item interactions are relatively harder to fit into a recommender model and tend to exhibit lower similarity score, and therefore assign lower or zero weight to these noisy interactions (i.e., $w_{ui}\propto s_{ui}$) to mitigate the adverse effects. 

\subsection{Empirical Studies of Denoised MMRecs}
\label{empiricalstudies}
In this section, we conduct experiments to explore whether the aforementioned denoising methods are still suitable for MMRecs. Specifically, we directly adapt representative denoising methods (T-CE \cite{wang2021denoising}, R-CE \cite{wang2021denoising}, and WBPR \cite{gantner2012personalized}) to representative MMRecs (VBPR \cite{he2016vbpr} and BM3 \cite{zhou2023bootstrap}) and plot their training curves in Figure~\ref{trainingcurve}. Based on the results, we have the following observations:
\begin{itemize} [leftmargin=8pt, topsep = 1pt]
    \item When equipped with denoising losses, MMRecs take more epochs to converge, and the training curves show a longer ``stagnation'' period where the performances do not increase for a few epochs. This phenomenon indicates that directly applying denoising loss functions to MMRec might provide incorrect denoising signals that actually confuses the training of MMRecs.
    \item MMRecs equipped with denoising methods do not perform well, sometimes even performing worse than normal training. Such a phenomenon indicates that traditional denoising methods originally designed for the IDRecs might not be suitable for MMRecs.
\end{itemize}

To further investigate this phenomenon, we randomly add 10\% noisy user-item interactions to the training data as ``oracle'' noises to examine whether the denoising methods can detect the oracle noises under the IDRec and MMRecs settings. Since they rely on the similarity score $s_{ui}$ to determine noisy interactions, we separately train an IDRec (LightGCN~\cite{he2020lightgcn}) and an MMRec (BM3~\cite{zhou2023bootstrap}) on the aforementioned noisy data and see if there is a difference between the $s_{ui}$ distributions of clean and the noisy interactions. As clean interactions far outnumber noisy interactions, we randomly sampled 10\% of clean interactions together with all noisy interactions and plotted their similarity scores in Figure~\ref{scoredistribution}. 

We can see that the difference in $s_{ui}$ distribution between the clean and noisy interactions on IDRec is relatively discernible, while on MMRecs the $s_{ui}$ distributions of clean and noisy interactions almost overlap. This phenomenon indicates that, \emph{as the representation space of IDRec and MMRecs is very different, the similarity score is no longer an accurate indicator of noisy interactions in MMRecs}. While these denoising methods can successfully detect noisy interactions in the IDRec setting, they are no longer suitable for MMRecs.

\section{Methodology}
\begin{figure}
    \centering
    \includegraphics[width=\linewidth]{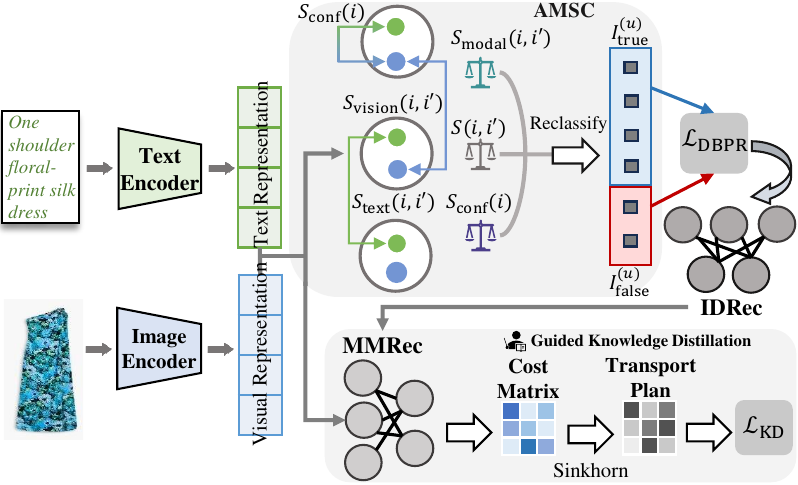}
    \caption{Framework overview of GUIDER.}
    \label{overview}
\end{figure}

In this section, we introduce the proposed GUIDER and elaborate on how we address different types of noise problems encountered in MMRecs, as shown in Figure \ref{overview}. Specifically, our GUIDER consists of three main parts: identifying noisy interactions from Adaptive Modality Similarity Calibration (AMSC), denoising user feedback, and guided knowledge distillation. We summarize the training algorithm of GUIDER in Algorithm \ref{GUIDER}.

\subsection{Identifying Noisy Interactions}
In MMRecs, features from multiple modalities are integrated to provide users with personalized recommendations. However, noisy modal content, such as irrelevant images or text descriptions, can make it more difficult to discern true user-item interactions from false ones \cite{zhang2017deep,wei2019mmgcn}. To address this challenge, we introduce Adaptive Modality Similarity Calibration (AMSC), a novel approach that aims to identify noisy interactions with fine-grained re-calibration from multi-modal semantics, thereby effectively identifying noisy interactions with the help of multi-modality features.

We initially sort the interactions based on the user-item loss. We set the average loss as the threshold to divide the low-loss and high-loss interactions into reliable samples $I^{(u)}_{\text{rel}}$ (i.e., likely to be clean interactions) and spurious samples $I^{(u)}_{\text{spr}}$(i.e., likely to be noisy interactions) respectively, which are calculated as follows:
\begin{equation}
\label{partition}
\begin{aligned}
    I^{(u)}_{\text{rel}} &= \big\{i \mid \mathcal{L_{\text{Rec}}}(u, i) \leq \frac{1}{|I_u|} \sum_{i \in I_u} \mathcal{L}(u, i), \forall i \in I^{(u)}\big\} \\
    I^{(u)}_{\text{spr}} &= \big\{i \mid \mathcal{L_{\text{Rec}}}(u, i) > \frac{1}{|I_u|} \sum_{i \in I_u} \mathcal{L}(u, i), \forall i \in I^{(u)}\big\}
\end{aligned}
\end{equation}
where $I^{(u)}=\{i|(u,i,r_{ui})\in\mathcal{D}\}$ denotes the set of items associated with user $u$, and $\mathcal{L}_{\text{Rec}}(u, i)$ (Eq. ~\ref{recommendationloss}) denotes the recommendation loss for the interaction between user $u$ and item $i$.

However, in MMRecs, such a partition might introduce false samples, as the loss value might not be an accurate indicator of the noise level. To mitigate this issue, we redistribute the reliable and spurious samples by leveraging the intra-modal semantic similarity among items. Our approach is based on the idea that while an item interacted with by a user may be distant from an uninteracted item in the recommendation semantic space, they may exhibit proximity in terms of modality. By integrating intra-modal signals to identify these samples, we can reduce the risk of mistakenly identifying real clean interactions as noise and improve the robustness of our denoising framework. Additionally, we introduce the inter-modal semantic matching of the current item as confidence scores to adjust the distribution of positive and negative samples.

Specifically, we calculate the modality similarity $S(i, i^{'})$ between each item $i$ in $I^{(u)}_{\text{spr}}$ and each item $i^{'}$ in $I^{(u)}_{\text{rel}}$. If $S(i, i^{'})$ is greater than the modality similarity threshold $S_{\text{thres}}$, we reclassify $i$ as a clean user-item interaction $I^{(u)}_{\text{true}}$, and obtain the noisy user-item interactions $I^{(u)}_{\text{false}}=I^{(u)}_{\text{rel}}\cup I^{(u)}_{\text{spr}}\setminus I^{(u)}_{\text{true}}$ accordingly. Our calibration function is formulated as follows:
\begin{equation}
\label{reclassify}
\begin{aligned}
    &i \in I^{(u)}_{\text{true}},\quad if\\ 
    &i\in I^{(u)}_{\text{rel}} \quad or \quad \exists i^{'}, i\in I^{(u)}_{\text{spr}}\land i^{'}\in I^{(u)}_{\text{rel}}\land S(i, i^{'}) >    S_{\text{thres}}
\end{aligned}
\end{equation}
where $S(i, i^{'})$ is calculated by integrating the intra-modality similarity $S_{\text{modal}}$ and the confidence level $S_{\text{conf}}$:
\begin{equation}
\label{modalitysimilarity}
    S(i, i^{'}) = S_{\text{modal}}(i, i^{'}) \times S_{\text{conf}}(i)
\end{equation}
Let $\mathbf{T}_i$ and $\mathbf{V}_i$ denote the text and vision representations of item $i$ extracted from modal-specific encoders, the intra-modality similarity $S_{\text{modal}}$ is calculated as the maximum of text modality similarity and the vision modality similarity:
\begin{equation}
    \begin{aligned}
        &S_{\text{modal}}(i, i^{'}) = \max(S_{\text{text}}(i, i^{'}), S_{\text{vision}}(i, i^{'})) \\
        &S_{\text{text}}(i, i^{'}) = \frac{\mathbf{T}_i \cdot \mathbf{T}_i}{\|\mathbf{T}_i\|\cdot\|\mathbf{T}_i\|}, \quad S_{\text{vision}}(i, i^{'}) = \frac{\mathbf{V}_i \cdot \mathbf{V}_{i^{'}}}{\|\mathbf{V}_i\|\cdot\|\mathbf{V}_{i^{'}}\|}
    \end{aligned}
\end{equation}

Aside from the intra-modality similarity, we believe that items whose representations are consistent across different modalities are more likely to be positive samples. High confidence suggests an accurate representation of consistent modalities, while low confidence indicates significant discrepancies between image and text descriptions. Therefore, we calculate the confidence level as follows:
\begin{equation}
\begin{aligned}
    &S_{\text{conf}}(i) = \text{CMH}(\mathbf{T}_i, \mathbf{V}_i) = \frac{\text{hash}(\mathbf{T}_i) \cdot \text{hash}(\mathbf{V}_i)}{||\text{hash}(\mathbf{T}_i)||\cdot||\text{hash}(\mathbf{V}_i)||}\\
    & \text{hash}(\mathbf{x}) = \text{sign}(\mathbf{W}^{\top} \mathbf{x} + \mathbf{b})
\end{aligned}
\end{equation}
where Cross-Modal Hashing (CMH) \cite{zhen2012co} is a technique that maps multi-modal representations into a common hash space to enable accurate alignment. The locality-sensitive hash function $\text{hash}(\mathbf{x})$ projects multi-modal data onto a hash space where similar items are likely to converge to preserve the inter-modal similarity signal. 

\subsection{Denoising User Feedback}
As we have mentioned, observed user-item interactions often fail to accurately reflect true user preferences, as they can be influenced by various factors, such as accidental clicks, misleading recommendations, or temporary curiosity, introducing significant noise into the collected feedback data \cite{Sun2019, Rendle2020,He2016, hu2008collaborative}. To address the challenge of noisy user feedback in recommendation systems, we propose an enhanced Denoising Bayesian Personalized Ranking (DBPR) loss function. 
The DBPR performs denoising from the positive and noisy negatives identified by AMSC, thereby enhancing the model's discrimination between genuine preferences and random noise.
For each user $u$, $\mathcal{L}_{\text{DBPR}}$ takes the clean set $I^{(u)}_{\text{true}}$ as positive samples and the noisy set $I^{(u)}_{\text{false}}$ as negative samples:
\begin{equation}
\label{DBPR}
 \mathcal{L}_{\text{DBPR}} = - \sum_{u \in \mathcal{U}, i \in I^{(u)}_{\text{true}}, j \in I^{(u)}_{\text{false}}} \log(\sigma(s_{ui} - s_{uj}))
\end{equation}
where $\sigma$ denotes the sigmoid function.
\subsection{Guided Knowledge Distillation}

As our preliminary experiments indicate that adopting denoising loss functions directly on MMRecs is infeasible, we bypass this issue by training a denoised IDRec as the teacher model to guide MMRec as the student model via Knowledge Distillation (KD) \cite{hinton2015distilling}, thereby effectively applying denoising techniques in MMRecs. The loss function of the student MMRec is defined as:
\begin{equation}\label{student}
\mathcal{L}_{\text{student}} = \mathcal{L}_{\text{KD}} + \mathcal{L}_{\text{Rec}}
\end{equation}
where $\mathcal{L}_{\text{KD}}$ is the KD loss which measures the discrepancies between the student and teacher model outputs. 

While the standard KL-divergence for KD may suffer from over-smoothing \cite{kim2016sequence} and model-collapse \cite{arjovsky2017towards}, we propose a novel Optimal Transport (OT) distance \cite{cuturi2013sinkhorn} based KD objective for distilling recommendation logits, which is more suitable as a symmetric distance metric for KD with stable gradient \cite{shen2018wasserstein}. The OT distance \cite{cuturi2013sinkhorn}, integrating Wasserstein distance with entropy regularization, provides a robust framework for aligning distributions with minimal overlap. It measures the cost of transporting mass between distributions, capturing the geometric structure and providing a robust metric even when distributions are disjoint. The OT-based KD loss function is formulated as the Frobenius inner product between the optimal transport plan $\mathbf{P}_\lambda$ and the cost matrix $\mathbf{D}$:
\begin{equation}
\mathcal{L}_{\text{KD}} = \langle \mathbf{P}_\lambda, \mathbf{D} \rangle
\end{equation}
The cost matrix \(\mathbf{D}\) is computed as the $L_2$ distance between the teacher logits distribution $\mathbf{z}^t$ and the student logits distribution $\mathbf{z}^s$:
\begin{equation}
\mathbf{D}_{m,n} = \left\| \mathbf{z}^t_m - \mathbf{z}^s_n \right\|^2_2
\end{equation}
where the logits distribution is computed based on the pairwise ranking score $\mathbf{z}=\{\log(\sigma(s_{ui} - s_{uj}))|u\in\mathcal{U}\}$ (Eq.\ref{DBPR}), superscript $t, s$ denote whether the logits come from the teacher or the student respectively, subscript $m, n$ denote the $m$-th and $n$-th element.

The optimal transport plan $\mathbf{P}_\lambda$ is obtained by solving optimal transport problem regularized by the entropy $H(\mathbf{P})$:
\begin{equation}
\mathbf{P}_\lambda = \mathop{\arg\min}_{\mathbf{P} \in U(\mathbf{z}^t, \mathbf{z}^s)} \sum_{x,y} \mathbf{P}_{x,y} \mathbf{D}_{x,y} - \lambda H(\mathbf{P})
\end{equation}
where $H(\mathbf{P}) = -\sum_{x,y} \mathbf{P}_{x,y} \log \mathbf{P}_{x,y}$ is the entropy of the transport plan $\mathbf{P}$, $U(\mathbf{z}^t, \mathbf{z}^s)$ denotes the set of all transportation plans that transform the teacher logits distribution \(\mathbf{z}^t\) into the student logits distribution \(\mathbf{z}^s\). This set includes all matrices \(\mathbf{P} \in \mathbb{R}_+^{d \times d}\) that satisfy the following marginal constraints:
\begin{equation}
U(\mathbf{z}^t, \mathbf{z}^s) = \left\{ \mathbf{P} \in \mathbb{R}_+^{d \times d} \mid \mathbf{P} \mathbf{1}_d = \mathbf{z}^s, \; \mathbf{P}^\top \mathbf{1}_d = \mathbf{z}^t \right\}
\end{equation}
where \(\mathbf{1}_d\) is a \(d\)-dimensional vector of ones. 

We adopt the Sinkhorn algorithm \cite{cuturi2013sinkhorn} to solve the the optimal transport plan $\mathbf{P}_\lambda$, which is calculated through iterative updates with the normalizing vectors $\mathbf{u}^l$ and $\mathbf{v}^l$:
\begin{equation}
\left(\mathbf{u}^l,\mathbf{v}^l\right)\leftarrow \left(\mathbf{z}^{s}\oslash\left(\mathbf{K}^{\top}\mathbf{v}^{l-1}\right),\mathbf{z}^{s}\oslash\left(\mathbf{K}\mathbf{u}^{l-1}\right)\right)
\end{equation}
where $\oslash$ denotes the element-wise division operator. The kernel matrix $\mathbf{K}$ with entropy regularization $\lambda$, which controls the smoothness of the optimization landscape and balances the transport cost and entropy in the OT distance computation, is computed as follows:
\begin{equation}
\mathbf{K} = \exp\left(-\mathbf{D}/{\lambda}\right)
\end{equation}
Upon convergence, the transport plan matrix $\mathbf{P}_\lambda$ is given by:
\begin{equation}
\mathbf{P}_\lambda = \text{diag}(\mathbf{u}^{l}) \mathbf{K} \text{diag}(\mathbf{v}^{l})
\end{equation}
where $\text{diag}(\cdot)$ denotes the diagonal matrix, which scales the rows and columns of \(\mathbf{K}\) to satisfy the marginal constraints.

\begin{algorithm}
\caption{The Training Procedure of GUIDER}
\label{GUIDER}
\small
\KwIn{$\mathcal{D}$: the user-item interaction data, $\mathcal{L}_{\text{Rec}}$: the recommendation loss, $S_{\text{thres}}$: the threshold for reclassification, an IDRec teacher $T$, an MMRec student $S$}
\KwOut{a well-trained MMRec $S$}

  \While{not converged}{
    Update teacher $T$ with the denoised loss function (Eq.\ref{DBPR})
    
    \For{each user $u$ in $\mathcal{U}$}{
        Divide samples into $I^{(u)}_{\text{rel}}$, $I^{(u)}_{\text{spr}}$ with $\mathcal{L}_{\text{Rec}}$ (Eq.~\ref{partition})
        
        Repartition $I^{(u)}_{\text{rel}}$ , $I^{(u)}_{\text{spr}}$ as $I^{(u)}_{\text{true}}$, $I^{(u)}_{\text{false}}$ with Eq.~\ref{reclassify}
    }
    Update student $S$ with the KD loss function (Eq.\ref{student})
}
\end{algorithm}
\section{Experiments}
\subsection{Experimental Settings}
\subsubsection{\textbf{Datasets.}}
We employ three multi-modal datasets from the Amazon platform \cite{mcauley2015image}, namely, Baby, Sports, Clothing. These datasets comprise rich textual and visual features in the form of item descriptions and images. Additionally, we utilize a multi-modal dataset from a short video platform, MicroLens \cite{ni2023content}. Compared to the Amazon dataset, the short video content from MicroLens features more complex interactions between video covers and textual descriptions within the context of social media platforms, introducing additional noise and complexity \cite{ni2023content}. After preprocessing according to the evaluation pipeline in \cite{zhou2023comprehensive}, we report dataset statistics in Table \ref{tab:dataset}.
\begin{table}
  \centering
  \small

    \caption{Dataset statistics after preprocessing.}
    \begin{tabular}{lcccc}
    \toprule
    Dataset & \# Users & \# Items & \# Interactions & Sparsity \\\midrule
    Baby  & 19,445 & 7,050 & 160,792 & 99.88\% \\
    Sports & 35,598 & 18,357 & 296,337 & 99.95\% \\
    Clothing & 39,387 & 23,033 & 237,488 & 99.97\% \\
    MicroLens& 98,129& 17,228& 705,174& 99.97\%\\
    \bottomrule
    \end{tabular}%
  \label{tab:dataset}%
\end{table}%
\subsubsection{\textbf{Metrics.}}
We employ the Recall and Normalized Discounted Cumulative Gain (NDCG) metrics at two cutoffs: $k=5$ and $k=20$. To ensure a fair comparison, we follow the same evaluation protocol of \cite{zhou2023bootstrap,zhou2023comprehensive} with a data splitting ratio of 8: 1: 1 on the interaction history
of each user, and rank the items in the whole dataset.
\subsubsection{\textbf{Baselines.}}
We select four representative MMRecs as base model to test different denoising methods:
\begin{itemize}[leftmargin=8pt, topsep=1pt]
    \item \textbf{VBPR} \cite{he2016vbpr} is a matrix factorization-based model that extends traditional collaborative filtering by incorporating visual features.
    \item \textbf{MMGCN} \cite{wei2019mmgcn} is a graph neural network model multiple types of item features for micro-video recommendation.
    \item \textbf{BM3} \cite{zhou2023bootstrap} is another self-supervised learning approach, aiming to enhance recommendation accuracy through multi-modal data.
    \item \textbf{DRAGON} \cite{zhou2023enhancing} constructs homogeneous graphs for MMRec to capture the complex relationships from user-item interactions.
    \item \textbf{FREEDOM} 
    \cite{zhou2023tale} simultaneously freezes the item-item graph and denoises the user-item interaction graph for MMrec.
\end{itemize}

\noindent Each base model is trained with GUIDER and the following state-of-the-art denoising baselines for comparisons:
\begin{itemize}[leftmargin=8pt, topsep=1pt]
    \item \textbf{WBPR} \cite{gantner2012personalized} considers items that are popular yet not interacted with by the user as true negative feedback, utilizing item interaction frequency to assess popularity and filter noise.
    \item \textbf{Truncated Cross-Entropy (T-CE) Loss} \cite{wang2021denoising} adjusts the loss function to truncate the contribution of ``hard'' interactions, which are identified by their loss values exceeding a threshold, thus pruning potentially noisy data during model training.
    \item \textbf{Reweighted Cross-Entropy (R-CE) Loss} \cite{wang2021denoising} applies a weighting function to the loss values, reducing the influence of interactions deemed ``hard'' by assigning them lower weights.
    \item \textbf{Bi-level Optimization for Denoising (BOD)} \cite{wang2023efficient} proposes an efficient bi-level optimization approach via one-step gradient matching to reweight the noisy implicit feedback.

\end{itemize}
\subsubsection{\textbf{Implementation Details}}
In line with best practices \cite{mcauley2015image, ni2023content}, all models use an embedding size of 64 for users and items. Embeddings are initialized with the Xavier method \cite{glorot2010understanding}. Optimization is performed using the AdamW optimizer \cite{kingma2014adam} with a learning rate searched within [1e-4, 1e-3]. Early stopping is employed to prevent overfitting, halting training if no improvement is observed on the validation set for 10 consecutive epochs. The weight decay for $L_2$ regularization is tuned within [1e-4, 1e-2]. For generality, we adopt LightGCN \cite{he2020lightgcn} as the default IDRec teacher.

\begin{table*}[t]
\centering
\caption{Performance comparisons of different denoising methods on multi-modal recommendation datasets. * indicates performance improvements that are statistically significant with p-value < 0.05.}
\label{tab:overall-performance}
\resizebox{\textwidth}{!}{%
\begin{tabular}{c|cccc|cccc|cccc|cccc}
\toprule
\textbf{Dataset} & \multicolumn{4}{c|}{\textbf{Baby}} & \multicolumn{4}{c|}{\textbf{Sports}} & \multicolumn{4}{c|}{\textbf{Clothing}} & \multicolumn{4}{c}{\textbf{MicroLens}} \\\midrule
 \textbf{Model}& R@5 & R@20 & N@5 & N@20 & R@5 & R@20 & N@5 & N@20 & R@5 & R@20 & N@5 & N@20 & R@5 & R@20 & N@5 & N@20 \\\midrule

VBPR&0.0265 & 0.0747& 0.0170& 0.0284&  0.0353& 0.0856& 0.0235& 0.0384& 0.0186& 0.0612& 0.0124&  0.0165& 0.0333& 0.0818& 0.0213& 0.0351\\
+T-CE& 0.0222& 0.0658& 0.0147& 0.0224& 0.0200& 0.0486& 0.0128& 0.0214& 0.0147& 0.0596& 0.0101& 0.0155& 0.0297& 0.0799& 0.0201&0.0345\\
+R-CE& 0.0264& 0.0680& 0.0172& 0.0292& 0.0248& 0.0617& 0.0164& 0.0270& 0.0150& 0.579& 0.0102& 0.0160& 0.0256& 0.0779& 0.0179&0.0322\\
+WBPR& 0.0234& 0.0568& 0.0154& 0.0230& 0.0202& 0.0501& 0.0136& 0.0220& 0.0144& 0.0587& 0.0094& 0.0156& 0.0327& 0.0789& 0.0200&0.0341\\
+BOD& \underline{0.0294}& \underline{0.0759}& \underline{0.0181}& \underline{0.0301}& \underline{0.0392}& \underline{0.0917}& \underline{0.0267}& \underline{0.0410}& \underline{0.0209}& \underline{0.0674}& \underline{0.0137}& \underline{0.0199}& \underline{0.0342}& \underline{0.0863}& \underline{0.0207}& \underline{0.0375}\\
 +GUIDER-DBPR& 0.0271& 0.0749& 0.0175& 0.0296& 0.0364& 0.0869& 0.0243& 0.0391& 0.0193& 0.0622& 0.0137& 0.0174& 0.0337& 0.0825& 0.0215&0.0360\\
+GUIDER & \textbf{0.0309*}& \textbf{0.0763*}& \textbf{0.0197*}&  \textbf{0.0329*}& \textbf{0.0415*}& \textbf{0.0983*}& \textbf{0.0275*}& \textbf{0.0421*}& \textbf{0.0232*}& \textbf{0.0797*}& \textbf{0.0154*}&  \textbf{0.0246*}& \textbf{0.0356*}& \textbf{0.0944*}& \textbf{0.0226*}& \textbf{0.0394*} \\

\midrule
MGCN & \underline{0.0365}& 0.0952& 0.0238& \underline{0.0412}& 0.0442& \underline{0.1088}& \underline{0.0290}& \underline{0.0491}& \underline{0.0390}& \underline{0.0939}& \underline{0.0256}& \underline{0.0418}& \underline{0.0402}& \underline{0.1050}& \underline{0.0247}& \underline{0.0441}\\
 +
T-CE& 0.0316& 0.0923& 0.0225& 0.0359& 0.0430& 0.1070& 0.0251& 0.0447&  0.0372& 0.0931& 0.0230& 0.0411& 0.0335& 0.0921& 0.0228&0.0419\\
 +R-CE& 0.0362& 0.0955& 0.0243& 0.0368& 0.0378& 0.0933& 0.0249& 0.0439& 0.0341& 0.0925& 0.0247& 0.0417& 0.0398& 0.1024& 0.0216&0.0414\\
 +WBPR& 0.0361& 0.0952& 0.0244& 0.0364& 0.0335& 0.0726& 0.0223&  0.0347& 0.0356& 0.0937& 0.0233& 0.0408& 0.0403& 0.1038& 0.0234&0.0408\\
+BOD& 0.0357& \underline{0.0971}& \underline{0.0242}& 0.0417& \underline{0.0447}& 0.1082& 0.0277& 0.0489& 0.0385& 0.0941& 0.0255& 0.0420& 0.0402& 0.1049& 0.0247& 0.0437\\
 +GUIDER-DBPR& 0.0364& 0.0951& 0.0231& 0.0410& 0.0438& 0.1076&  0.0281& 0.0479& 0.0381& 0.0932& 0.0251& 0.0417& 0.0395& 0.1043& 0.0241&0.0438\\
+GUIDER & \textbf{0.0384*}& \textbf{0.0997*}& \textbf{0.0253*}& \textbf{0.0428*}& \textbf{0.0457*}& \textbf{0.1116*}& \textbf{0.0302*}& \textbf{0.0506*}& \textbf{0.0403*}& \textbf{0.0959*}& \textbf{0.0268*}& \textbf{0.0430*}& \textbf{0.0416*}& \textbf{0.1084*}& \textbf{0.0263*}& \textbf{0.0457*}\\

\midrule

BM3 & 0.0327& 0.0883 & 0.0216& 0.0383 & 0.0353& 0.0980& 0.0238& 0.0438& 0.0246& 0.0641&  0.0164& 0.0294& 0.0300& 0.0857& 0.0189& 0.0347\\
 +T-CE& 0.0277& 0.0793& 0.0160& 0.0279& 0.0318& 0.0919& 0.0220& 0.0419
& 0.0207
& 0.0611
& 0.0124
& 0.0247
& 0.0278
& 0.0845
& 0.0137
&0.0309
\\
 +R-CE& 0.0321& 0.0823& 0.0207& 0.0355& 0.0307& 0.0907& 0.0256& 0.0385& 0.0173
& 0.0577
& 0.0073
& 0.0234
& 0.0215
& 0.0838
& 0.0113&0.0268
\\
 +WBPR& 0.0317&  0.0841& 0.0201& 0.0353& 0.0308& 0.0910& 0.0226& 0.0377& 0.0189
& 0.0575
& 0.0099& 0.0227
& 0.0241
& 0.0825& 0.0124
&0.0281
\\
+BOD& \underline{0.0341}& \underline{0.0911}& \underline{0.0224}& \underline{0.0397}& \underline{0.0372}& \underline{0.1004}& \underline{0.0259}& \underline{0.0441}& \underline{0.0277}& \underline{0.0689}& \underline{0.0183}& \underline{0.0326}& \underline{0.0325}& \underline{0.0881}& \underline{0.0202}& \underline{0.0366}\\
+GUIDER-DBPR&0.0329 &0.0891 &0.0218 &0.0389 &0.0361 &0.0985 &0.0242 &0.0440 &0.0253 &0.0652 &0.0171 &0.0301 &0.0310 &0.0864 &0.0193 &0.0355\\
+GUIDER  & \textbf{0.0380*}& \textbf{0.0981*}& \textbf{0.0248*}& \textbf{0.0419*}& \textbf{0.0430*}& \textbf{0.1047*}& \textbf{0.0283*}& \textbf{0.0458*}& \textbf{0.0393*}& \textbf{0.0853*}& \textbf{0.0259*}& \textbf{0.0391*}& \textbf{0.0364*}& \textbf{0.0965*}& \textbf{0.0231*}& \textbf{0.0430*}\\ 
\midrule

 DRAGON& \underline{0.0366}& \underline{0.1021}& \underline{0.0239}& \underline{0.0425}& \underline{0.0439}& \underline{0.1100}& \underline{0.0296}& \underline{0.0495} & \underline{0.0392}& \underline{0.0947}& \underline{0.0260}& \underline{0.0417}& \underline{0.0465} &\underline{0.1096} & \underline{0.0299}&\underline{0.0487}\\
  +T-CE& 0.0320
&   0.1016
& 0.0213
& 0.0418
& 0.0401& 0.1097& 0.0287& 0.0488& 0.0382& 0.0943& 0.0259
& 0.0388
& 0.0451
& 0.1088
& 0.0254
&0.0471
\\
+R-CE& 0.0257
& 0.0882
& 0.0162
& 0.0345
& 0.0354
& 0.0906
& 0.0255
& 0.0404
& 0.0334
& 0.0823
& 0.0169
& 0.0324
& 0.0371
& 0.0896
& 0.0207
& 0.0376
\\
+WBPR& 0.0281
& 0.0878
& 0.0155
& 0.0365
& 0.0318
& 0.0884
& 0.0249
& 0.0406
& 0.0332
& 0.0816
& 0.0182
& 0.0344
& 0.0372
& 0.0910
& 0.0225
& 0.0440
\\
+BOD& 0.0359& 0.0997& 0.0233& 0.0413& 0.0427& 0.1086& 0.0272& 0.0485& 0.0379& 0.0932& 0.0254& 0.0407& 0.0441& 0.1103& 0.0274& 0.0468
\\
 +GUIDER-DBPR&0.0355 &0.1003 &0.0232 &0.0415 &0.0422 &0.1088 & 0.0284& 0.0483&0.0382 &0.0931 &0.0246 &0.0403 &0.0457 &0.1082 &0.0288 &0.0471\\
+GUIDER & \textbf{0.0385*}&\textbf{0.1137*}& \textbf{0.0254*}& \textbf{0.0447*}& \textbf{0.0465*}& \textbf{0.1129*}& \textbf{0.0310*}& \textbf{0.0507*}& \textbf{0.0404*}& \textbf{0.0980*}& \textbf{0.0277*}& \textbf{0.0436*}& \textbf{0.0507*}& \textbf{0.1142*}& \textbf{0.0327*}&\textbf{0.0511*}\\

 \midrule

FREEDOM & \underline{0.0367}& \underline{0.0992}& \underline{0.0236}& \underline{0.0424} & \underline{0.0438}& \underline{0.1082}& \underline{0.0288}& \underline{0.0481} & \underline{0.0389}& \underline{0.0941}& \underline{0.0263}& \underline{0.0420}& \underline{0.0451} &\underline{0.1094} & \underline{0.0272}&\underline{0.0471}\\
  +T-CE& 0.0304
& 0.0964
& 0.0201
& 0.0406
& 0.0391& 0.1057& 0.0274& 0.0468& 0.0362& 0.0899& 0.0246
& 0.0369
& 0.0426
& 0.1029
& 0.0239
&0.0446
\\
 +R-CE& 0.0271
& 0.0962
& 0.0177
& 0.0371
& 0.0381
& 0.0975& 0.0276& 0.0437& 0.0366& 0.0891& 0.0183& 0.0348
& 0.0403
& 0.0971
& 0.0227
&0.0409
\\
 +WBPR& 0.0306
& 0.0976
& 0.0167
& 0.0397
& 0.0345& 0.0967& 0.0269& 0.0442& 0.0357& 0.0900& 0.0196& 0.0371
& 0.0411
& 0.1010
& 0.0242&0.0467
\\
+BOD& 0.0338
& 0.0987
& 0.0218
& 0.0416
& 0.0394
& 0.1075
& 0.0277
& 0.0468
& 0.0365
& 0.0916
& 0.0239
& 0.0373
& 0.0439
& 0.1043
& 0.0253
& 0.0452
\\
+GUIDER-DBPR &0.0358 &0.0981 &0.0224 &0.0407 &0.0437 &0.1076 &0.0277 &0.0474 &0.0372 &0.0926 &0.0255 &0.0408 &0.0438 &0.1085 &0.0264  &0.0469\\   
 +GUIDER&\textbf{ 0.0386*}&\textbf{ 0.1052*}& \textbf{0.0249*}& \textbf{0.0446*}& \textbf{0.0462*}& \textbf{0.1119*}& \textbf{0.0301*}& \textbf{0.0498*}& \textbf{0.0403*}& \textbf{0.0972*}& \textbf{0.0270*}& \textbf{0.0430*}& \textbf{0.0492*}& \textbf{0.1127*}& \textbf{0.0302*}&\textbf{0.0496*}\\
 \midrule
 \end{tabular}
}
\end{table*}

\subsection{Overall Performance}
\label{overallperformance}
We compare GUIDER with other denoising frameworks on four base multi-modal models across four datasets. For comparison, we also report a variant GUIDER-DBPR, which is trained with the knowledge distillation from teacher only without DBPR. From the results in Table \ref{tab:overall-performance}, we have the following observations:

\textbf{Effectiveness of Denoising:} Our method exhibits clear advantages across various datasets, notably in short-video scenarios that face higher noise levels. For example, when combined with BM3, GUIDER achieves a significant increase in Recall@5 on MicroLens. Due to MicroLens being relatively noisier compared to other datasets, GUIDER is able to achieve a larger margin on MicroLens, indicating its effectiveness in recommendation denoising.

\textbf{Different Impact Across Various Base Models:} The comparison among different base models under our framework reveals that the effectiveness of GUIDER is remarkably enhanced when applied to complex base models capable of handling detailed, multi-modal interactions. For example, from the MicroLens dataset, we observe sophisticated models like BM3 and DRAGON benefit significantly from GUIDER. This showcases our framework's ability to better adapt to more cutting-edge MMRecs, which generally get more complex with the development of recommendation technology. 

\textbf{Superior Denoising in Multi-Modal Frameworks:} Unlike traditional denoising methods, our approach excels in complex multi-modal environments through AMSC, significantly enhancing the ability to discern noises from various modalities. Traditional denoising methods like T-CE, R-CE, and WBPR adopt a one-size-fit-all approach, which do not fully exploit the fine-grained interactions between feedbacks and multi-modal contents, resulting in marginal improvements in the MMRecs setting. Even advanced framework like BOD also struggles in multi-modal settings due to their limited capacity to manage the nuanced interactions between different data modalities. In contrast, GUIDER is tailored for MMRecs and can not only reduce noise but also mitigate the inconsistency between the ID and the modality representation spaces, leading to substantial gains in performances compared with baseline denoising methods.

\textbf{Limitations of Using KD Alone:} While one may wonder the effectiveness of GUIDER might come from KD, we empirically show that using KD alone without DBPR (GUIDER-DBPR) does not lead to performance improvement or only marginal improvement. Since the teacher model is trained without denoising, it remains susceptible to data noise and consequently cannot effectively transfer useful knowledge to the student. As a result, the student MMRec sees minimal to no performance gains from KD alone. It is only when KD is complemented with DBPR that MMRec can truly benefit from the denoising effect introduced by DBPR.

\subsection{Ablation Study}
\label{sec:ablation}
We employ BM3 as the base model and conduct an ablation study on the following components:
\begin{itemize}[leftmargin=*]
    \item \emph{w/o KD}: This variant disables knowledge distillation (KD) and directly applies AMSC and DBPR to MMRec.
    \item \emph{w/o DBPR}: This variant disables the DBPR component, which also causes AMSC to be ineffective, leaving only the effect of KD.
    \item \emph{w/o AMSC}: This variant disables the AMSC component, reverting to a standard method of selecting positive and negative samples.
    \item \emph{w/ KL Divergence}: This variant uses the standard KL divergence for KD instead of the OT distance.
\end{itemize}
\begin{table}
    \caption{Ablation study on the key components of GUIDER.}
    \centering
    \small
    \setlength{\tabcolsep}{1pt}
    \begin{tabular}{c|cc|cc|cc|cc}
        \hline
        Data & \multicolumn{2}{c|}{Baby} & \multicolumn{2}{c|}{Sports} & \multicolumn{2}{c|}{Clothing} & \multicolumn{2}{c}{MicroLens} \\
        \hline
        Metrics & R@20 & N@20 & R@20 & N@20 & R@20 & N@20 & R@20 & N@20 \\
        \hline
        \hline
        \emph{w/o} KD & 0.0875 & 0.0391 & 0.0997 & 0.0442 & 0.0616 & 0.0285 & 0.0815 & 0.0327 \\
        \emph{w/o} DBPR & 0.0891 & 0.0389 & 0.0985 & 0.0443 & 0.0652 & 0.0301 & 0.0864 & 0.0355 \\
        \emph{w/o} AMSC & 0.0912 & 0.0407 & 0.1011 & 0.0453 & 0.0706 & 0.0329 & 0.0882 & 0.0365 \\
        \emph{w/} KL Div. & 0.0953 & 0.0419 & 0.1035 & 0.0464 & 0.0807 & 0.0373 & 0.0932 & 0.0408 \\
        \hline 
        GUIDER & \textbf{0.0981} & \textbf{0.0419} & \textbf{0.1047} & \textbf{0.0458} & \textbf{0.0853} & \textbf{0.0391} & \textbf{0.0965} & \textbf{0.0430} \\
        \hline
    \end{tabular}
    \label{tab:ablation_results}
\end{table}
Based on the results in Table~\ref{tab:ablation_results}, we can see that:

(1) w/o KD: Directly applying AMSC and DBPR to MMRec without knowledge distillation results in minimal improvement or even a slight decrease in performance. This highlights the challenge of directly applying denoising loss functions to MMRec due to the complexity of multi-modal data.

(2) w/o DBPR: Using KD alone without DBPR does not lead to significant performance improvements because the teacher model is still affected by noisy interactions (Section \ref{overallperformance}).

(3) w/o AMSC: The absence of AMSC results in a notable decrease in performance. This indicates that AMSC plays a crucial role in aligning multi-modal content with user preferences by handling inconsistencies in multi-modal representation.

(4) w/ KL Div.: Compared to the OT distance, the KL divergence suffers from over-smoothing and model collapse. Additionally, KL divergence fails to effectively utilize ID information to guide MMRec, as it cannot capture complex distributional differences between teacher and student models as accurately as the OT distance.
\subsection{Hyperparameter Sensitivity}
\subsubsection{\textbf{Effects of Regularization Coefficient}}

\begin{figure}[ht]
    \centering
    \includegraphics[width=0.48\textwidth]{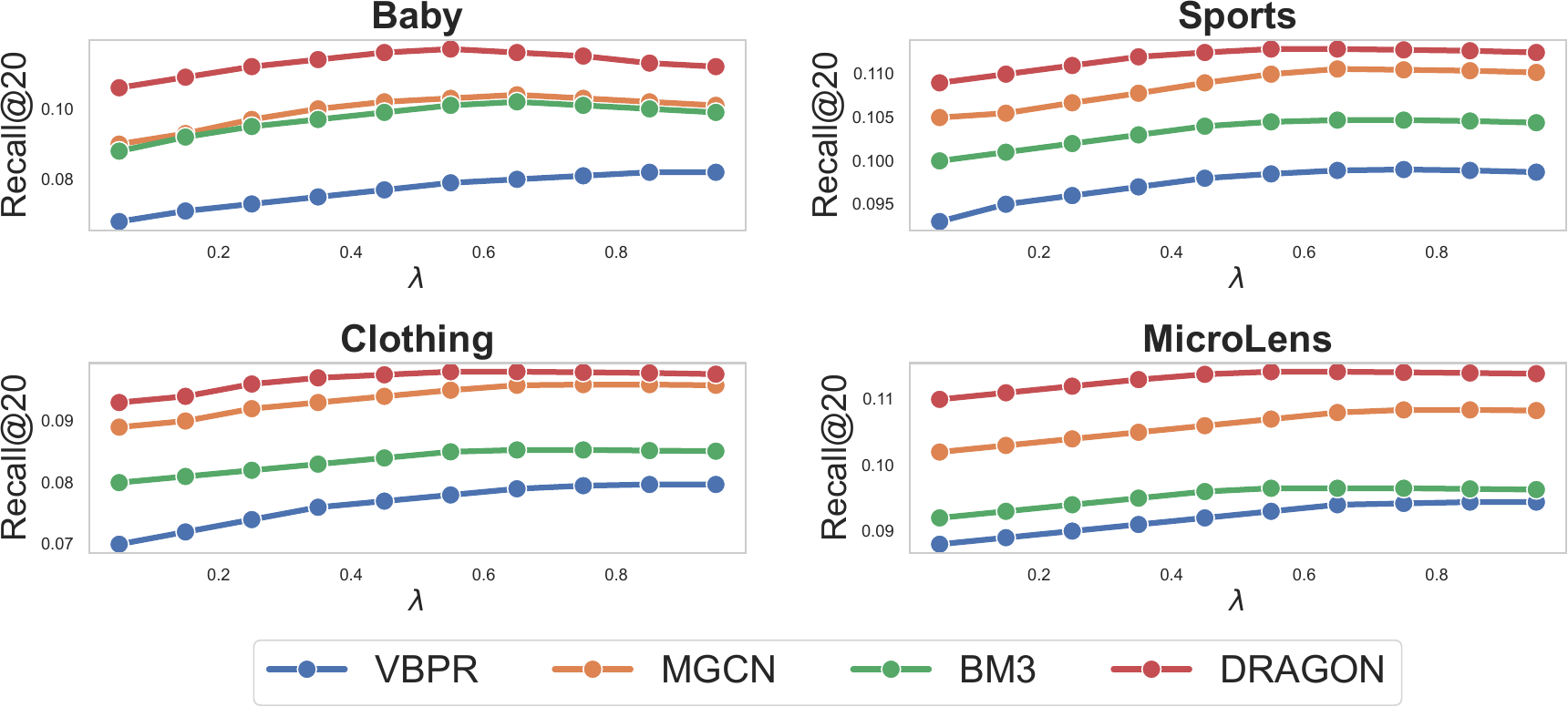}
    \caption{Analysis of $\lambda$ sensitivity in KD.}
    \label{fig:lambda}
\end{figure}

The regularization coefficient $\lambda$ in the Sinkhorn algorithm plays a crucial role in controlling the trade-off between the fidelity to the original cost and the entropy of the transport plan. To evaluate the sensitivity of the regularization coefficient $\lambda$, we plot the performances of MMRecs on four datasets under different values of $\lambda$ in Figure~\ref{fig:lambda}. Based on the results, we have the following observations.

\begin{itemize}[leftmargin=8pt,topsep=1pt]
    \item The GUIDER framework demonstrates a commendable level of stability across a range of $\lambda$ settings, suggesting a well-calibrated response to the regularization parameter. The results indicate that even as the impact of entropy regularization varies, the efficacy of our framework can also be ensured.
    
    \item We find that models with lower baseline performance benefit more significantly from higher $\lambda$ values during KD. This trend supports the notion that by increasing the influence of entropy, higher $\lambda$ values can effectively enrich the training signals for these models and thus enhance the learning effectiveness.
\end{itemize}


\subsubsection{\textbf{Threshold Sensitivity in AMSC}}

\begin{figure}[ht]
    \centering
    \includegraphics[width=0.48\textwidth]{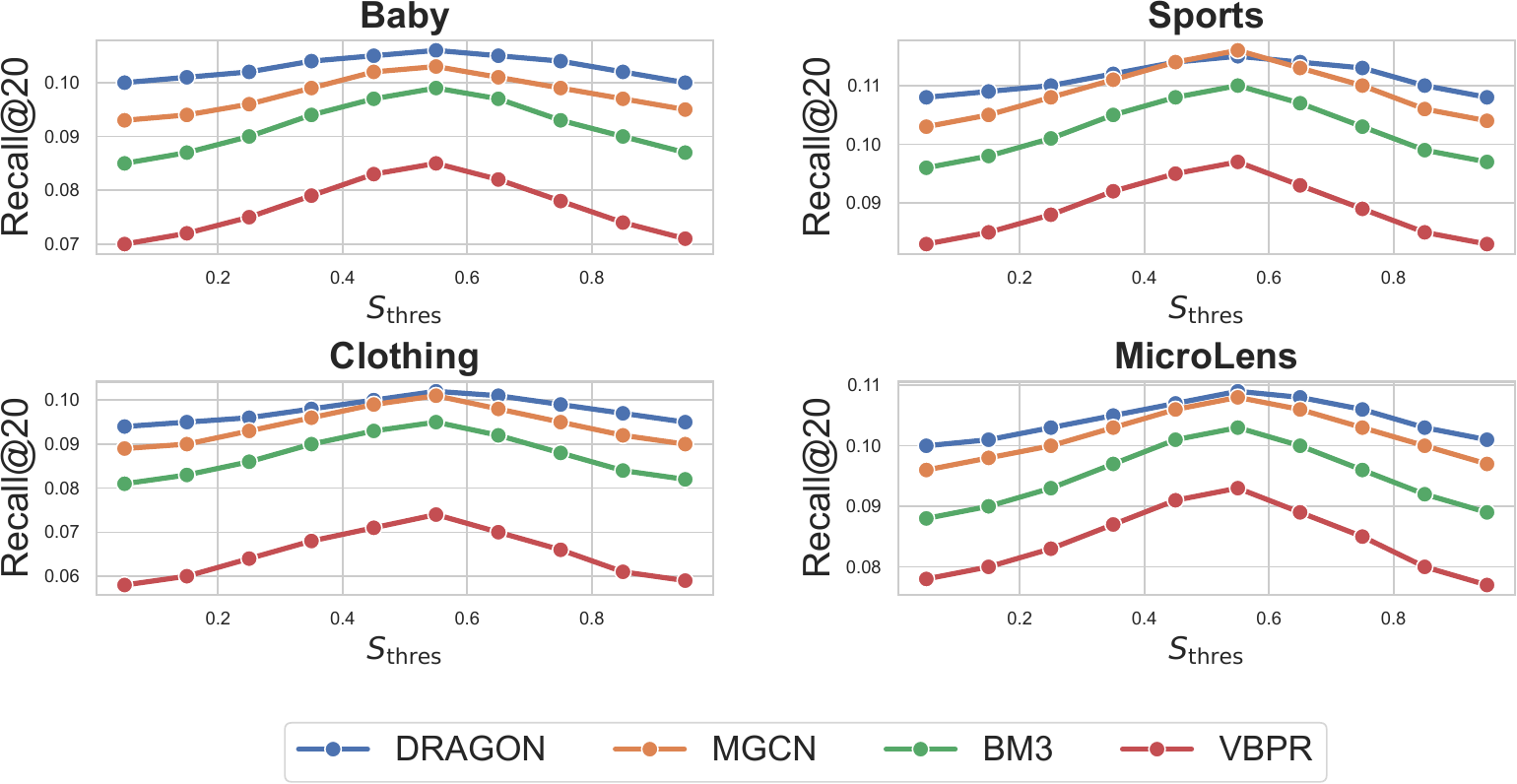}  
    \caption{Analysis of $S_{\text{thres}}$ sensitivity in AMSC.}
    \label{fig:threshold_sensitivity}
\end{figure}

The threshold hyperparameter \( S_{\text{thres}} \) within the AMSC component delicately balances the modulation of false positive samples against true positive samples. Based on the results in Figure~\ref{fig:threshold_sensitivity}, we have the following observations:

\begin{itemize}[leftmargin=*]
    \item GUIDER's denoising performance remains relatively stable with threshold alterations, suggesting that within a reasonable range, the precise value of the threshold is not critically sensitive.
    \item At a lower threshold of 0.7, GUIDER shows propensity to incorrectly classify a substantial number of false positives as true recommendations, resulting in sub-optimal performance. 
    \item Conversely, setting the threshold to 1 substantially undermines the denoising capability of the AMSC. At this point, the absence of AMSC’s discernment leads to a deterioration in performance, reflecting the component's value in handling intra- and inter-modality balance for better recommendation quality.
\end{itemize}

\subsection{Robustness in Noisy Scenarios}

\begin{figure}[htbp]
\centering
\includegraphics[width=\linewidth]{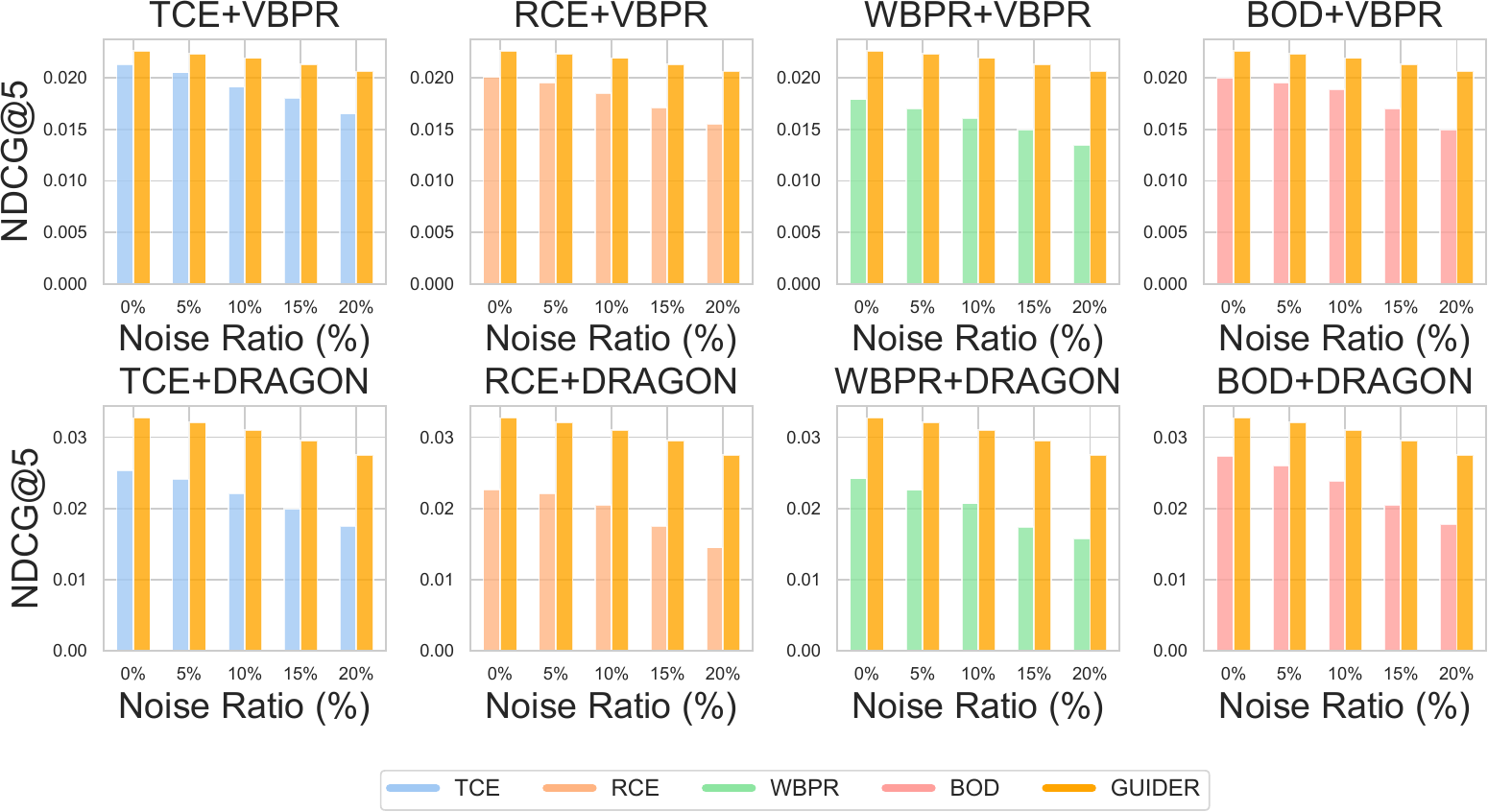}
\caption{Performances with increasing noise ratios.
}
\label{fig_overall}
\end{figure}

In exploring the robustness of various denoising methods, we observe that while traditional techniques struggle to effectively suppress noise as the noise level increases, GUIDER maintains superior performance. To demonstrate GUIDER's robustness under severe noises, we randomly add a certain proportion (5\%, 10\%, 15\%, 20\%) of noise to the original training data from the MicroLens dataset and examine the performance of GUIDER under these noisy settings. From the results in Figure~\ref{fig_overall}, we have the following observations:

\begin{itemize}[leftmargin=*]
\item Traditional MMRec models experience a significant performance decline as the noise ratio increases, highlighting the detrimental effect of noise on recommendation accuracy. The deterioration becomes more significant with higher noise ratios, illustrating the sensitivity of mainstream MMRec to noise interference.
\item The incorporation of the GUIDER framework not only enhances the performance of these models across all noise levels but also markedly mitigates the rate of performance decline, demonstrating GUIDER's dual capability to boost overall recommendation effectiveness in noisy scenarios while concurrently stabilizing the models against the disruptive influence of noise.
\end{itemize}

\section{Conclusion}

We introduce GUIDER, a universal framework for enhancing denoising in multi-modal recommender systems. GUIDER leverages an AMSC strategy for fine-grained recalibration of multi-modal semantics, incorporates a DBPR loss to filter noisy user feedback, and addresses modal representation inconsistencies via guided distillation from teacher to student models. Nonetheless, there are a few limitations: the AMSC module may encounter computational complexity challenges on large-scale datasets, potentially impacting real-world efficiency. Further, future work could enhance GUIDER’s robustness to noise and explore its adaptability across diverse data types and recommendation settings.

\bibliographystyle{ACM-Reference-Format}
\balance

\end{document}